# A relativistic time variation of matter/space fits both local and cosmic data

Alfredo Gouveia Oliveira[*] and Rodrigo de Abreu[†]

[*]*Direcção Técnica, RDP, Lisboa, Portugal*
[†]*Departamento de Física and Centro de Electrodinâmica, IST, Lisboa, Portugal*
E-mail: alfredooliveira@rdp.pt and rabreu@ist.utl.pt

Cosmic observations strongly support a time varying scenario for matter/space. On the other hand, so far, observations at solar system scale failed to identify any time variation on matter/space characteristics. To explain both results it is enough to consider a time variation of physical parameters liable to account for cosmic observations but satisfying Relativity Principle at least at local scale (we called it Local Relativity Property). Here, for the first time, a relativistic time varying scenario is defined from redshift and Cosmic Microwave Background characteristics. It is showed that it can match both cosmic and local data. Although undetectable in each local measure, such relativistic scenario has detectable time dependent consequences at Solar System scale, namely a receding component in the motion of the Moon, a past warmer climate and, this being new, an accelerating component in Earth rotation. A new class of cosmological models can now be explored, without concepts like dark matter, valid at both local and cosmic scale, and such that fundamental physical laws and Relativity Principle hold in any point of time and space.

## I. INTRODUCTION

In this work we begin by investigating a set of four relevant results from experience concerning fundamental properties of Nature. We do not question the validity of those results, namely we do not question whether Michelson-Morley experiment has or has not an exact null result - we accept the standard results to the first approximation. Those results are the following:
  a) Misfit between cosmic data and physical laws - the locally discovered fundamental physical laws, established considering phenomena in the neighborhood of the observer, i.e., considering only local data, do not directly fit cosmic data;
  b) Relativity property - physical laws seem to be valid whatever the inertial translatory motion of a reference frame where the measure of the one-way light speed has the value of the circulation speed of light. We designate such reference frame by Einstein reference frame because it is carefully defined by Einstein in Special Relativity [1]. It is commonly designated Lorentz reference frame because it supports Lorentz coordinates transformation. The generalization to whatever motion is not analyzed in this work;
  c) Constancy of the circulation speed of light, i.e., of the average light speed in an elementary closed path;
  d) Light speed independence in relation to the motion of its source.

The last three results represent the fundamental properties of Nature on which Special Relativity (SR) is based. The large number of experiments that have been made to test SR are mainly tests of these three results. The first result represents a property of Nature discovered after SR. Therefore, it was not considered by Einstein in his work of 1905.

In a previous paper [2], we presented the core of the analysis, the framework in it established being now applied to cosmic data. In the following, we summarize the results obtained in that paper, beginning with the fundamental statements of our analysis.

The first of the four results from experience, i.e., the misfit between cosmic data and physical laws, can be basically a consequence of one of two possibilities: either there is a phenomenon unknown at local scale that explains the misfit or current physical laws are not valid at a cosmological scale, being only the local limit of more general physical laws.

It is not the first time that a misfit between cosmic data and physical laws is observed and one can identify the above alternatives as the Ptolemy's approach and Galileo's approach. Ptolemy's approach is more economic, it does not question existing theories, while Galileo's approach implies rewriting some part of physics. We are, of course, simplifying a complex question, in order to introduce a crucial decision of our analysis: to consider that physical laws are locally valid but to do not presume their non-local validity. This decision, however, is not a consequence of the above considerations but of the simple fact that we want to be very rigorous in our analysis. Physical laws relate measures made by an atomic observer (our instruments) relative to the same time moment and concerning distances that are small compared with cosmic distances, i.e., they just fit snapshots of phenomena in our neighborhood. Hence, and once they do not directly fit cosmic observations, there is no support from experience to presume their non-local validity.

The first statement of this analysis is, therefore, that fundamental physical laws are just **local physical laws**, i.e., *the limit of general physical laws when r→0 and dt→0 in an Einstein reference frame*.

A consequence of attributing only local validity to physical laws is that to interpret cosmic data one needs now to find out something on how phenomena depend on position, in space and in time. The analysis of the range data to the Viking landers on Mars [3, 4], intended to detect at solar system scale consequences of a space expansion or of a time variation of fundamental constants, gave a negative result. Although relative to a small range in space and time, that result can plausibly support the statement that *local physical laws are valid in any point of space-time*.

Considering this, the second result from experience can then be stated as: *local physical laws are independent of reference frame motion or position in space and time*. This statement is a restriction of Relativity property to snapshots of phenomena in the neighborhood of the observer and a generalization of it to position satisfying Cosmological Principle. We named it the **Local Relativity Property** (LRP). It enables us to link local and cosmic data. Note that LRP statement must be understood as valid "to the first approximation" and not necessarily in "any" point but only in a large range of space-time.

Relativity Principle can now be understood as the statement that phenomena can be described by general physical laws that are a function only of measures relative to the observer. The difference between LRP and Relativity Principle is that LRP is local. We assume that LRP is valid but one has to find out whether or not Relativity Principle, which applies to non-local phenomena, holds.

The third result, the constancy of the circulation speed of light, can support a clear definition of proper time unit, which we designate simply by time unit because it is with it that our perception of time is built: **Time Unit** is *such that the circulation speed of light is $c_0$*. Note that, differently of length unit, time unit is a scalar.

Finally, light speed source independence means that in each point of the Universe light velocity in empty space may eventually depend on the position of the point in time-space, on the local gravitational field, on the overall distribution of matter, on some unknown characteristic of the Universe, but does not depend on the motion of its source. This is a critical result because, if the propagation of fields is related with light propagation, then one can expect that matter properties depend on motion. If light/field propagation was "ballistic", one could expect that matter/space characteristics were independent of motion; however, this does not seem to be the case. In part I we show that Sagnac effect can be explained considering only the constancy of light circulation speed and light speed source independence, being an evidence from experience of these two properties.

A necessary consequence (we are not considering that spacetime has any peculiar property because, until now, that is not a necessary consequence of any result from experience) of the two last results is that velocity measuring unit has to change with the motion of the reference frame so that the measure of light circulation speed may keep constant. This conclusion is very easy to obtain, impossible to ignore, and implies a directional change of $LT^{-1}$, i.e., a change with motion of length measuring unit $L$ and, eventually but not necessarily, a change also of time unit $T$.

An important aspect one must note is that one cannot measure a change in matter characteristics by the analysis of a local atom with null velocity in relation to the observer. The reason is that the measuring units change in the same way as matter properties. On the other hand, one can detect such a change by the analysis of systems with a motion, in a field or with a position in time and space different from the one of the observer. For instance, if the mass or the shape of an atom depends on motion, one cannot detect it by directly measuring the mass or shape of a local atom with null relative velocity but one may, eventually, be able to detect that the same atom with a non-null relative velocity has a different mass or shape; or if atomic properties change with time, one cannot detect it locally but one may, eventually, to detect it by analyzing atoms on a different time position.

The above reasoning applies directly only to fundamental magnitudes; the general analysis of physical systems properties, represented by physical laws, is a little more complex. It is shown, in part I, that matter/space changes are undetectable by local measures, in an Einstein reference frame, provided that the variation of fundamental physical parameters obeys a set of three conditions, called Local Relativity conditions. These conditions are just the dimensional equations of fundamental constants.

A common presumption that is not a necessary consequence of experience is that an atomic observer is invariant. This is supported on the apparent invariance of the perception of local phenomena. It is important to understand that such invariance only implies Local Relativity Property and LRP does not require atomic invariance, only a variation of physical parameters in accordance with LR conditions and the use of an Einstein reference frame, as referred above.

Another common presumption not supported in experience is that the one-way velocity of light is constant. It is important to understand that Einstein postulate of the constancy of light speed concerns a property of the reference frame and the constancy of light circulation speed but not the one-way velocity of light. Einstein was careful but eventually not clear enough on his postulate of light speed constancy; what the postulate states (in other words) is that light speed is constant in the velocity measurements made in a reference frame where Newton laws are valid. Relativity property holds only in such reference frame. In Einstein deduction of coordinates transformation it is clearly considered that it is only the measure of light speed in the reference frame that has to be constant, this imposing the time coordinates so that such result is obtained. This contradiction with a somewhat simplistic idea that one-way light speed relative to the observer is constant disregarding the reference frame has been favoring Minkowski approach, Einstein analysis being frequently ignored in Relativity courses. The constancy of one-way light speed is a property of the reference frame, not of light, although being only possible because of the constancy of light circulation speed. In part I we show that *the synchronization of clocks presented by Einstein cannot support any meaningful measure of one-way light speed because, whatever this one, the result of the measure is always light circulation speed.* The fact that one cannot interpret Einstein postulate of the constancy of light speed as representing a physical property of one-way light speed in relation to the observer has been fully analyzed by some authors [5, 6].

Note that the above does not change the mathematics of Special Relativity, as presented by Einstein. The two postulates of Special Relativity correctly represent relativity property, an Einstein reference frame being characterized by the constancy of the ratio of the variation of time and length coordinates along a light path. This is not arbitrary but physically meaningful: the velocity of phenomena depend on light/field velocity in relation to source and receiver and an appropriate time unit vector can be defined as $c_0$ times the time interval spent by light in a length unit path. One can call this the coordinate time unit and corresponds to the time unit referred by Einstein in his postulate of light speed constancy, i.e., it is the time unit used for calibrating the reference frame where Newton laws hold.

Because atomic properties can (eventually) vary with motion, field or position, two atomic observers with different motions, or in different fields, or in different points of space or time, can present measurable differences. In order to conveniently analyze the variation of matter/space properties, we introduced, in part I, the concept of invariant observer, such as two invariant observers present no differences whatever each one motion, field or position.

To finish the definition of the framework of Local Relativity, it is yet necessary to find out whether or not the four results from experience imply a connection between space and time. In part I, in a 3-Euclidean space with no connection between time and space, using the invariant observer and the $LT^{-1}$ dependence with velocity obtained from the constancy of the circulation speed of light, we concluded that, in an Einstein reference frame, Doppler effect, aberration laws (at least in a simple case) and Maxwell laws are independent of the uniform translatory motion of the reference frame. We characterized an Einstein reference frame in Euclidean space and obtained that the coordinates transformation between two Einstein reference frames with uniform translatory motion is Lorentz transformation, affected by a function $\varphi$ that Einstein concluded to be $\varphi=1$, based in the presumption that $\varphi$ is a function of reference frames relative velocity. As our analysis of Doppler effect, aberration laws and Maxwell laws concluded that these results from experience are independent of $\varphi$, we find no evidence from experience that supports the above Einstein conclusion. This is an important result because the possibility of being $\varphi \neq 1$ has deep consequences in the understanding of this entire subject. Part I ends with the analysis of mathematical and physical reasons of Time and Length paradoxes associated with Lorentz transformation.

Special Relativity corresponds locally to Local Relativity in atomic metric but Local Relativity fully clarifies SR concepts of timespace and relative simultaneity as well as time and length paradoxes of Lorentz transformation.

In part I we established the framework that characterizes the invariance of Local Physical Laws, called Local Relativity. This framework defines how physical parameters can vary so that local physical laws hold. Experience seems to show, this being stated by Local Relativity Property, that local physical laws hold whatever the inertial motion, gravitational field or position of the Einstein reference frame. Therefore, the second phase of the analysis is to determine the variation of physical parameters in each one of these cases, i.e., the variation of physical parameters implied by the invariance of local physical laws in spite of the motion, field or position of the observer.

In this paper we initiate the analysis of the positional case. Here, we begin by retrieving from Part I the essential information for positional analysis, concerning observers, notation, reference frames and LR conditions. Then, we define a relativistic time varying scenario from redshift and CMB (cosmic microwave background). Classical cosmic tests and local consequences are analyzed.

## II. THE FRAMEWORK OF THE POSITIONAL ANALYSIS

In the analysis here presented we disregard the influence of gravitational field on physical parameters.

### A. From Part I: observers and LR conditions

We use the two observers defined in part I: the *A* observer, which is such that a local atom with null relative velocity is invariant in relation to him, whatever his motion and position; and the *R* observer, which is such that the measuring units of two *R* observers are always identical, whatever their relative motion, field or position. The fundamental units of measure used are Mass, Charge,

Length and Time. A superscript identifies the observer to whom the unit belongs and a subscript identifies the observer that measured the parameter. So, the fundamental units of measure of $A$ are $M^A$, $Q^A$, $L^A$ and $T^A$; $[G]^A$ represents the $A$ measuring unit of $G$; $G_A$ is the value of gravitational constant measured by $A$; $M_R^A$ represents the value of the mass unit of $A$ measured by $R$ ($M_R^A \equiv M^A/M^R$); $[G]_R^A \equiv [G]^A/[G]^R$ is the value of the $A$ unit of $G$ measured by $R$.

The measuring units of $R$ and $A$ are equal at the point in time and space chosen as origin, named "zero point", considering an absence of field and that both observers are "at absolute rest", as defined in part I. The local measures made in this situation are identical whatever the observer, being identified by the subscript "0". LRP implies that the $A$ local value of physical parameters is always the same as the value measured in this situation, for instance, $G_A = G_0$.

Note that although we use two metrics, as in the LNH of Dirac [7], both metrics of Dirac correspond to our atomic units, the gauge function of scale-covariant cosmology [8] being constant in our framework.

As explained in part I, the invariance of the perception of local phenomena with null mean velocity by an atomic observer is satisfied by the invariance of his determination of the value of physical constants by local experiments. This implies that physical constants can only vary accordingly with their dimensional equations, which is expressed by LR conditions:

$$T = c^{-1}L$$
$$GM = c^2 L \qquad (1)$$
$$\sqrt{G/\varepsilon}\, Q = c^2 L$$

The letters in LR conditions (1) represent the relative value of magnitudes and constants in relation to their value in the "zero point", being the ratio between $A$ and $R$ measuring units, for instance $M = M_R^A$, or, being the same, the ratio between $R$ and $A$ measures, for instance, $c = c_R/c_A = c_R/c_0 = L_R^A/T_R^A$.

As measures are inversely proportional to measuring units, representing a generic physical entity by $phy$, it is:

$$phy_R = phy_A \cdot [phy]_R^A \qquad (2)$$

Since LRP implies the invariance of the $A$ measure of any physical entity, i.e., it implies $phy_A = constant$, then, from (2), the variation of $phy_R$ is equal to the variation (in $R$) of the $A$ measuring units of $phy$, $[phy]_R^A$. In other words, physical entities have to vary accordingly with their dimensional equations. For instance, the wavelength (in $R$) of a spectral radiation has to vary accordingly to $L$ because $[\lambda]_R^A = L$. Therefore, in case of a $L$ variation with time, the radiation of the same spectral line emitted in different time instants have different wavelengths, corresponding to the $L$ variation, so that a local $A$ observer will always measure the same value for the wavelength (at the emission moment). Also, the energy of radiation, at the emission moment, has to vary (in $R$) with energy dimension $ML^2T^{-2} = Mc^2$, the Planck constant with $McL$. Only in this way can phenomena keep invariant to $A$, as stated by the LRP.

When an $A$ observer makes non-local observations, the variation (in $R$) of non-local phenomena and of $A$ measuring units are no longer equivalent, therefore $A$ can detect a relative difference. For instance, in case of a $L$ variation with time, $A$ will detect a wavelength difference between a radiation from a distant source and the correspondent local radiation, this considering that there is no change in the wavelength from the source to $A$. On the other hand, dimensionless numbers, like the fine structure constant, have naturally to keep invariant in $R$ and, therefore, in $A$ non-local observations (considering that there is no change in light properties in the path between source and observer). The *invariance of dimensionless local combinations of physical parameters* can also be used as an enunciation of LRP. Albrecht and Magueijo [9] have already referred that physical experiments are only sensitive to dimensionless combinations of dimensional physical parameters. The non-local analysis of such numbers, namely the fine structure constant, at cosmic distances [10, 11, 12, 13], in different gravitational fields or at different velocities, are positional, field and dynamic tests for LRP and source of information on the change of light properties between the source and the observer.

B. On the Time and Space components of Position

Consider that light properties keep invariant between light source and the observer. Then, the observable consequence of a positional dependence of matter/space characteristics would be a non-local violation of local physical laws, e.g., the same phenomena in distant stars and in the Sun would present different characteristics to an Earth observer. A time dependence of matter/space would introduce an isotropic component in this violation and space dependence an anisotropic one. However, one cannot presume the invariance of light properties in the path source-observer. In case of space dependence of matter/space, the variation of light properties between source and observer could mask the source anisotropy, and the observer could only detect an isotropic component. Therefore, on interpreting cosmic observations one must be aware of the possible consequences a change in light properties in the path source-observer can have.

There are two fundamental observational phenomena that can be directly related with a positional variation of matter/space characteristics: redshift and cosmic microwave background (CMB). Both seem to be isotropic. Furthermore, isotropy seems to be a systematic characteristic of cosmic observations.

Bearing in mind all the above, *we will consider that the violations of local laws observed at cosmic scale result from a time dependence of physical parameters*. In what concerns spatial dependence, we will consider that if there is one with a relevant level within the cosmological horizon, it is completely masked by a change in light properties during light path and has no observational consequences. We will also consider that an earth observer can be adequately approximated by an observer "at absolute rest", disregarding gravitational field. Therefore, the one-way velocity of light in relation to him is isotropic whatever the definition for the time coordinate of the reference frame. The time position $t=0$ is present moment.

### C. Space Curvature

In this first analysis we consider that space is Euclidean to the $R$ observer because, at this point, there is no reason to presume otherwise. Accordingly with standard cosmologies, recent analyses of CMB anisotropies indicate a flat Universe [14]. Our analysis is limited to observable cosmic objects, which, in this case, differently of what happens in Friedmann models, are at a distance presumably much shorter than the distance to an antipodal point or some sort of frontier of the Universe; therefore, using an Euclidian space in this analysis does not imply any necessary incompatibility with a cosmological model with space curvature, namely a closed model.

## III. TIME DEPENDENCE OF PHYSICAL PARAMETERS FROM REDSHIFT AND CMB

In this chapter a time variation of physical parameters is defined after redshift and CMB properties and obeying LR conditions. With such a variation, physical laws keep fitting the local measures made in any time moment.

### A. Postulate of the redshift

A fundamental parameter of cosmic observations is redshift $z$. It relates the wavelength $\lambda_0$ of a local radiation and the wavelength $\lambda rec$ of a correspondent radiation received from a distant source, being $\lambda rec = \lambda_0(z+1)$. A precise observational result is the fitting of the cosmic microwave background (CMB) by the Planck formula for blackbody radiation at a temperature of $\approx 2.7$ °K. Designating by $dP(T,\lambda)$ the power flux in a $d\lambda$ interval given by Planck formula for temperature $T$ and wavelength $\lambda$ it is:

$$dP(T,\lambda) = (z+1)^4 \cdot dP[T/(z+1), \lambda \cdot (z+1)] \quad (3)$$

Therefore, a wavelength shift of $(z+1)$ transforms Planck formula for a temperature $T$ in Planck formula for temperature $T/(z+1)$ provided that an attenuation of $(z+1)^4$ also occurs. The phenomenon that produced the CMB is not known, although there is an explanation for it, but it is plausible to consider that whatever it may be, the radiation produced by it fits Planck formula for a local $A$ observer. In accordance with this, we will consider that *radiation received from distant sources presents a (z+1) wavelength shift and an attenuation of $(z+1)^4$ in relation to local radiation produced by similar phenomena*. We state this as a postulate because we cannot be sure, at this point, of the nature of the phenomenon that produced the CMB. We call it the **postulate of the redshift**. It implies that a radiation that fits Planck formula to a local atomic observer in some time instant also fits Planck formula for a commoving atomic observer along its time line, although for a time changing value of temperature.

### B. Time dependence of matter/space

It has already been noted that LRP implies that a variation of the wavelength $\lambda_R$ of spectral radiation has to be in accordance with $L$ variation. Therefore, the wavelength of a radiation emitted at a time $t$ is $\lambda_R(t) = \lambda_0 L(t)$, where $\lambda_0$ is today wavelength of the correspondent radiation. Between a distant source and us, the wavelength of the radiation can change if a change in light speed in $R$ occurs. This is a possibility we have to consider because the constancy of the local mean light speed (in $A$) does not imply the constancy of light speed in $R$. The problem a variation of vacuum light speed puts is to know what is the correspondent variation of wavelength. We have no observational evidence; however a wavelength variation (to $R$) in correspondence with $c_R$ would be in accordance with our local concepts. We will consider this instead of developing a general analysis because it allows a more clear analysis and still exact in the case of time invariance of light speed. This case is of crucial importance at this point because a first objective is to find out whether or not light speed is time invariant. Therefore, considering that (in $R$) *the wavelength of a traveling radiation keeps proportional to vacuum light speed*, the relation between the wavelength $\lambda_0$ of a local radiation and the wavelength $\lambda rec$ of a correspondent radiation received from a distant source is

$$\lambda rec = \lambda_0 \, c^{-1} L \quad (4)$$

As redshift $z$ is such that $\lambda rec = \lambda_0(z+1)$, then $z = c^{-1}L - 1$, where $c$ and $L$ are relative to the moment when the radiation was produced. A red shift is observed if $c^{-1}L>1$. Let us introduce a function of time $\alpha(t)$ defined as:

$$\alpha(t) = c^{-1}L = z+1 \qquad (5)$$

one obtains from LR conditions (1):

$$\begin{aligned} L &= c\alpha \\ T &= \alpha \\ GM &= c^3\alpha \\ \sqrt{G/\varepsilon}\, Q &= c^3\alpha \end{aligned} \qquad (6)$$

Any change through time of fundamental magnitudes and constants that obey conditions (6) satisfies the LRP and produces a redshift for a time decreasing $\alpha(t)$.

For both an atomic observer here and another one in a distant point in time and space, the Planck formula is verified in local phenomena, accordingly to LRP; as, from the postulate of redshift, to the observer here the radiation received also fits Planck formula, once the wavelength is shifted, the radiation has to suffer an attenuation of the power flux of $(z+1)^4=\alpha^4$ in the path between the distant observer and here, measured in $A$ units. As the $A$ measuring unit of power flux varies with $M\alpha^{-3}$, then the flux attenuation in $R$ is $M\alpha$. Therefore, the apparent attenuation in $A$ is partly or totally due to the variation of $A$ measuring unit. For $M = \alpha^{-1}$, i.e., a time increasing of mass, there is no attenuation in $R$; for $M = 1$, i.e., a time invariant mass, the attenuation in $R$ is $\alpha$; for $M = \alpha$, which can configure an evanescence of matter, the attenuation in $R$ is $\alpha^2$. In Appendix we detail the above results.

## IV. COSMOLOGICAL FRAMEWORK

The results of cosmic observations are usually expressed in terms of observational parameter $z$ and Friedmann models parameters $H_0$ and $q_0$. Friedmann models assume the Robertson-Walker line element, where the scale factor $S(t)$ is related with $z$ by $S(t)=S_0/(1+z)$. The Hubble constant is:

$$H_0 = \left(\frac{\dot{S}}{S}\right)_{t=0} = -\left(\frac{\dot{z}}{1+z}\right)_{t=0} \qquad (7)$$

As $\alpha = z+1$, $H_0$ can be defined as:

$$H_0 = -\left(\frac{1}{\alpha}\frac{d\alpha}{dt_A}\right)_{t=0} = -\left(\frac{d\alpha}{dt_R}\right)_{t=0} \qquad (8)$$

One procedure at this point could be to parameterize $\alpha(t)$ using a series development, for instance, using Friedmann parameters, as:

$$\alpha^{-1}(t_A) = 1 + H_0 t_A - \frac{q_0}{2}(H_0 t_A)^2 + O\!\left((H_0 t_A)^3\right) \qquad (9)$$

In case of a fit of cosmic observations with null $q_0$, one could conclude that $\alpha^{-1}(t_A)$ is a linear function; in case of a fit with constant $q_0$, one could conclude that $\alpha^{-1}(t_A)$ is not linear but one would not know $\alpha^{-1}(t_A)$ for large values of $z$ because the knowledge of the second order term, which is the maximum one can get from current cosmic observations, is not enough; in other cases, one could conclude that something is wrong with the theory. As the same conclusions can be obtained simply by considering a linear $\alpha^{-1}(t_A)$, there is no point in using a series development. On the contrary, a more clear analysis can be made considering a linear $\alpha^{-1}(t_A)$ instead of a series development. Therefore:

$$\alpha(t_A) = (1 + H_0 t_A)^{-1} \qquad (10)$$

An important result of (6) is that the time unit of $A$ depends only on $\alpha$, so (10) enables us to relate the measure of time in $A$ and in $R$:

$$t_R = \int_0^{t_A} T\, dt_A = \int_0^{t_A} \alpha\, dt_A = H_0^{-1} \ln(1 + H_0 t_A) \qquad (11)$$

and:

$$\begin{aligned} t_A &= -H_0^{-1}\frac{z}{z+1} \\ t_R &= -H_0^{-1} \ln(z+1) \\ \alpha(t_R) &= e^{-H_0 t_R} \end{aligned} \qquad (12)$$

Thus, $\alpha(t_R)$ is a negative exponential function of $t_R$. This has the advantage of not imposing an origin neither an end and has the self-similar behavior expected for the phenomenon, i.e., a time shift is equivalent to a scale change. The time dependence of physical parameters is then independent of absolute time, of an age of the Universe, and so are $R$ general physical laws. In this case, the time variation of matter/space obeys Relativity Principle. Therefore, this is a convenient and plausible test function for $\alpha$; however, only the results of observations, mainly at large $z$ or of large scale structure, can validate or not this choice.

We have also to parameterize $c$ or $L$. The prime objective at this point is to verify the possibility of the two most simple cases, $c$ invariant or $L$ invariant. To this one corresponds $c=\alpha^{-1}$, so a parameterization of $c$ as

$$c = \alpha^{-u} \quad (13)$$

can exactly represent those two cases, for $u=0$ and $u=1$. Note that the conclusions obtained from observations are only valid in the range of these ones, i.e., a conclusion that $u \approx 0$ obtained from observations at low $z$ do not imply an invariant light speed at high $z$.

## V. THE AGE OF MATTER AND OF STARS

The first of equations (12) shows that age calculations made in $A$ will tend to the value of $H_0^{-1}$ when relative to old cosmic objects or phenomena. Therefore, while matter can be much older in $R$ than the oldest stars, in $A$, stars will be almost as old as the matter itself. This is what happens with current calculations of the age of stars and of the Universe because they are calculations in $A$. This is a consequence of atomic time unit increase towards the past and can be considered an evidence of this increase. On the other hand, the time interval in $R$ from the appearance of matter until the first star being much larger then the time since that occurrence has important consequences in the analysis of large scale structure.

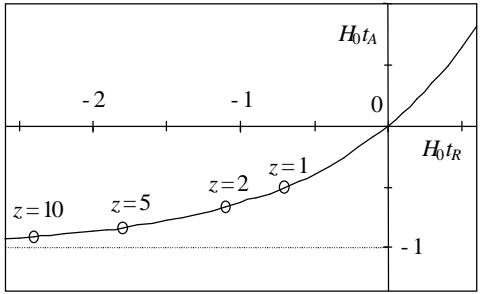

FIG. 1. The measure of time in $A$ and in $R$. Whatever the age of the matter (measured with the $R$ invariant time unit, in the abscissa) for an $A$ observer matter can be no older than $H_0^{-1}$. The time position of occurrences at $z=1$, $z=2$, $z=5$ and $z=10$ are shown.

## VI. COSMIC TESTS

Having defined the basic framework, we will now determine the equations for classic cosmic tests and compare them with the ones of Friedmann models with cosmological constant $\Lambda=0$, for simplicity. The first step is to determine the relations between distance, brightness and redshift. Consider a source with redshift $z$. The light now received was emitted at $t_R = -H_0^{-1}\ln(z+1)$. The distance $r_R$ to that source is ($c_R = c_0 \cdot c(t)$):

$$r_R = \int_{t_R}^{0} c_R dt_R = c_0 H_0^{-1} \ln(z+1) \cdot \bar{c} \quad (14)$$

where

$$\bar{c} = \frac{1}{t_R} \int_0^{t_R} c(t) dt_R \quad (15)$$

For $c$ invariant it is $r_R = c_0 H_0^{-1} \ln(z+1) = -c_0 t_R$ and for $L$ invariant it is $r_R = c_0 H_0^{-1} z/(z+1) = -c_0 t_A$.

The luminosity $\ell$ of a source, in the absence of evolution, has to be constant to $A$ for observation of the LRP; so, to $R$, because luminosity has the dimensions of power, it has to be:

$$\ell_R = \ell_0 M c^2 \alpha^{-1} \quad (16)$$

Considering (16), together with the distance-redshift relation (14) and the $R$ attenuation with $M\alpha$ of the radiation, one obtains for the flux $\Im_R$ received from the source:

$$\Im_R = \frac{\ell_0}{4\pi c_0^2} H_0^2 \left( \alpha \cdot \ln\alpha \cdot \frac{\bar{c}}{c} \right)^{-2} \quad (17)$$

Thus, the luminosity distance $D$ is:

$$D = \frac{c_0}{H_0} \alpha \cdot \ln\alpha \cdot \frac{\bar{c}}{c} \quad (18)$$

Note that, from (13) and (15):

$$\frac{\bar{c}}{c} = \frac{\alpha^u - 1}{u \ln\alpha} \quad (19)$$

For $c$ invariant ($u=0$) it is $\bar{c}/c = 1$ and for $L$ invariant ($u=1$) it is $\bar{c}/c = z/\ln(z+1)$.

### A. Distance-modulus / redshift.

From (18):

$$m - M = 42.38 - 5\log h_0 + 5\log(\alpha \cdot \ln\alpha) + 5\log(\bar{c}/c) \quad (20)$$

With an error less than $0.05m$, for $z<1$ and $|u|\leq 1$, the distance modulus can be approximated by:

$$m - M = 42.38 - 5\log h_0 + 5\log(z) + 2.5(1+u)\log(1+z) \quad (21)$$

Figure 2 displays the Hubble diagram for several values of $q_0$ and $u$. The possibility of an "accelerating" Universe, raised by the analysis of some supernova data [15], implies $u>0.2$. Other analyses [16] indicate a $q_0 \approx 0.1$, which corresponds to $u \approx 0$ for $z<1$.

### B. Time dilation

A necessary consequence of (6) is the dilation of phenomena with $A$ time unit. This is confirmed by a $(1+z)$ time dilation in the light curve of distant supernova [16, 17] and in spectral feature age measurements of a supernova [18].

### C. Angular-size / redshift

The diameter $d$ of a body varies with $L$ for an $R$ observer, i.e., $d_R = d_0 \alpha \cdot c$ where $d_0$ is the diameter for an $A$ observer or the diameter of an equivalent body in our neighborhood. Considering the distance/redshift expression (14), the angular-size/redshift is:

$$\theta = \frac{H_0}{c_0} d_0 \frac{z+1}{\ln(z+1)} \frac{c}{\bar{c}} \equiv \frac{H_0}{c_0} d_0 \alpha \frac{u}{\alpha^u - 1} \qquad (22)$$

Figure 2 displays $\theta(z)$ curves for several values of $u$ and $q_0$. For $u=0$ there is a minimum at $z=e-1=1.72$; for $u=1$ the angular-size tends to $d_0 H_0/c_0$.

Observational results indicate a $1/z$ law at low redshift, this being in accordance with both theories, and also at high redshift for double radio sources, a result that is not in accordance with neither theory. Compact radio sources show a different behavior. A recent work of Gurvits, Kellermann and Frey [19] based on a sample in the range $0.011 \le z \le 4.72$ obtained $q_0 = 0.21 \pm 0.30$ considering no evolutionary or selection effects. To the angular-size/redshift curve for $q_0 = 0.21$ corresponds $u = 0$ with an error of less than 3% in the $z$ range. To the above $q_0$ interval corresponds approximately $u = 0 \pm 0.5$.

### D. Surface brightness

From the magnitude and angular-size of a source one obtains:

$$\sigma = \frac{\ell_0}{\pi^2 d_0^2 (1+z)^4} \qquad (23)$$

This is independent of $\alpha(t)$ and $c(t)$ and depends only on $(1+z)$ as its negative fourth power, exactly as in the Robertson-Walker models.

### E. Number counts

The expression for the slope of the $\log N(m)$ curve, considering no evolution in the $A$ luminosity and in the $R$ number density of galaxies, is:

$$\frac{d \log N}{dm} = 0.6 \left( 1 + \frac{\bar{c}}{c} \ln \alpha \left( 1 - \frac{\alpha}{c} \frac{dc}{d\alpha} \right) \right)^{-1} \equiv 0.6 \left( \alpha^u + \frac{\alpha^u - 1}{u} \right)^{-1} \qquad (24)$$

For $u=0$:

$$\frac{d \log N}{dm} = \frac{0.6}{1 + \ln(1+z)} \qquad (25)$$

The slope for $u=0$ corresponds to $q_0=0$ with an error of less than 0.01 till $z=4$.

### F. Comparison with Friedmann models

The comparison with Friedmann models evidences identical behavior for surface-brightness test and time dilation and a close correspondence between $u=0$ and $q_0=0.1$ for the magnitude, $q_0=0.2$ for the angular-size and $q_0=0$ for number counts, in the observed range of $z$.

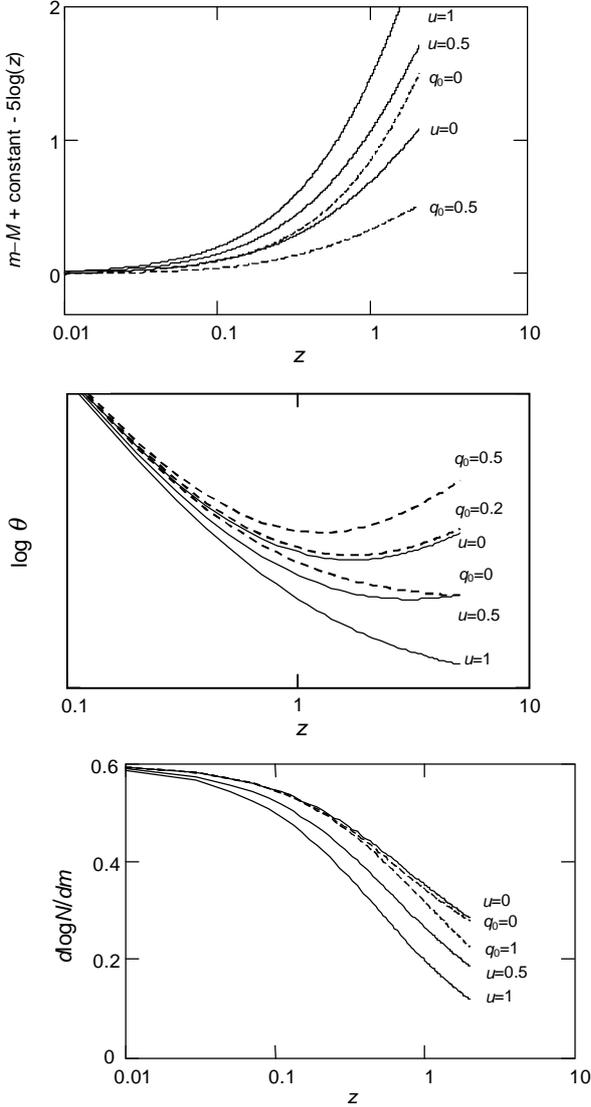

FIG. 2. The relations with redshift of magnitude, angular-size and number counts. Solid lines for Local Relativity and dashed ones for Friedmann models with $\Lambda = 0$.

## VII. SOLAR SYSTEM

### A. Generalized laws

To analyze the time dependence of solar system dynamics one needs $R$ physical laws with non-local validity in relation to time. One can obtain $R$ local laws from $A$ local laws using (3) to replace $A$ values of each physical entity *phy* by $R$ values. The general situation is that $[phy]_R^A$, which depends on velocity, field and position, vary with direction and, therefore, the laws obtained this way are complex. LRP is not valid in $R$ but only in relation to an atomic observer, i.e., in $A$. However, for a first analysis of solar system evolution one can disregard the consequences of velocity, field and space position (if there is any) on physical parameters. Then, $[phy]_R^A$ does not depend on direction. Now, one has also to consider that local physical laws can depend on the time derivative of some physical entity *phy*. From equation (3):

$$\frac{d(phy_R)}{dt_R} = \frac{d(phy_A)}{dt_A} \cdot [phy]_R^A \cdot \frac{dt_A}{dt_R} + (phy_A) \cdot \frac{d[phy]_R^A}{dt_R} \quad (26)$$

As $dt_A/dt_R = (T_R^A)^{-1}$ and $[phy]_R^A/T_R^A = [d(phy)/dt]_R^A$, equation (26) takes the form of equation (3) for $d[phy]_R^A/dt_R = 0$, which is satisfied by a dimensionless *phy*. Then, because they are dimensionally consistent, $R$ and $A$ local laws have the same form in an isotropic scenario if they contain no time derivatives or if time derivatives apply only on dimensionless combinations of parameters. We will designate these laws by *time generalized local laws*. Naturally, in $R_0$, the values of constants are the $R_0$ values and the results are in $R_0$ units, while in $A$ the values are $A$ values and the results are in $A$ units.

An example of a local law without time derivatives is Planck law. Therefore, the generalized Planck law corresponds to the usual one. Newton laws, on the contrary, are local laws with a time derivative and have to be transformed. Noting that

$$v_R = v_A \cdot [v]_R^A = \frac{v_A}{c_0} c_R \Rightarrow \frac{v_R}{c_R} = \frac{v_A}{c_A} \quad (27)$$

a way of formulating the generalized inertia law is:

$$\frac{d}{dt}\left(\frac{v}{c}\right) = 0 \quad (28)$$

The absence of a subscript in time generalized laws does not indicate the relative variation of a parameter but that the equation is valid both for $A$ and $R$ values in an isotropic scenario. The generalized second local law of Newton can then be formulated as:

$$F = mc\frac{d(v/c)}{dt} \quad (29)$$

The acceleration of the gravitational field produced by a mass $m$ can be expressed by:

$$\frac{d}{dt}\left(\frac{v}{c}\right) = \frac{G}{c}\frac{m}{r^2} \quad (30)$$

B. Conservation Laws

LRP applies only to phenomena of elementary duration. Therefore, it does not imply any time conservation, namely a time conservation of energy. On the contrary, in case of time variation of $A$ energy units, time conservation of energy cannot happen both in $A$ and in $R$. On the other hand, LRP does not interfere with momentum or energy conservation in phenomena with elementary duration. This is the case of linear momentum and energy conservation in elastic collision, or of angular momentum conservation in a motion ruled by a central force.

The time dependence of angular momentum is important for the analysis of orbital motion we present in the next point. In $R$, a definition of momentum liable to display time conservation, either linear or angular, must be a function of $v/c$ and cannot depend on fundamental magnitudes because these can vary with time. One can define an intensity angular momentum $J_R$ as:

$$\mathbf{J_R} = \frac{\mathbf{v_R}}{c_R} \times \mathbf{r_R} \quad (31)$$

As a central force does not change the value of $\mathbf{v} \times \mathbf{r}$, one can conclude that in the case of a motion where the only force is central it is $J_R = constant$.

C. Orbital motion

A simple analysis, using simple calculations, of some basic orbital characteristics can be made considering a local circular orbit, i.e., an orbital motion that tend to circular when $dt \to 0$. Note that, as referred, we are neglecting the motion and field dependence of physical parameters. In this case, the acceleration of the gravitational field produced by a mass $m$ can be expressed by the generalized law (30) and:

$$\dot{\mathbf{v}} = G\frac{m}{r^2} \cdot \frac{\mathbf{r}}{r} + \mathbf{v}\frac{\dot{c}}{c} \quad (32)$$

The last term of the equation has the direction of $\mathbf{v}$, so it does not contribute to the centripetal acceleration which, in a circular orbit, is $v^2/r$. Therefore, a first equation that characterizes a local circular orbit is $v^2 r = Gm$. A second equation is obtained considering that the conservation of the angular momentum in case of a central force implies that in this case of a local circular orbit it is, from equation (31), $J_R = constant$. Note that in case of a time varying light speed the force is not exactly central because the second component of the acceleration in (32) makes the curvature radius not coincident with $\mathbf{r}$. This is, however, just a high order effect, one of the several we are neglecting here.

Thus, noting that, from (6), $G_R m_R = G_0 m_0 \alpha c^3$, a local circular orbit is characterized by:

$$\begin{aligned} v_R^2 r_R &= G_R m_R = v_0^2 r_0 \alpha c^3 \\ \frac{v_R}{c_R} r_R &= \frac{v_0}{c_0} r_0 \end{aligned} \quad (33)$$

The solution is:

$$\begin{aligned} r_R &= r_0 \alpha^{-1} c^{-1} \\ v_R &= v_0 \alpha c^2 \end{aligned} \quad (34)$$

Therefore, the orbital radius varies inversely to the length measuring unit of $A$ ($L = \alpha c$). The measures in $A$ of orbital radius and angular velocity $\omega$ are

$$\begin{aligned} r_A &= r_0 \alpha^{-2} c^{-2} \\ \omega_A &= \omega_0 \alpha^3 c^3 \end{aligned} \quad (35)$$

Therefore, $\omega_A^2 r_A^3 = G_A m_A = constant$. This is of crucial importance. From his local measurements of the planetary system, an $A$ observer will always determine the same value for $G$, even if it varies in $R$. This is new and different from what happens in other theories that have considered an evolutionary Universe and have led to a search for the value of $\dot{G}/G$ [20, 21, 22].

To obtain the present ratio of variation of orbital characteristics it is not necessary to parameterize either $\alpha(t)$ or $c(t)$ because only the value of their $A$ time derivatives at $t = 0$ is needed. By definition, the $A$ time derivative at $t = 0$ of $\alpha$ is $(\dot{\alpha})_{A,0} = -H_0$. Making, without any loss of generality, $(\dot{c})_{A,0} = uH_0$ one obtains:

$$\begin{aligned} \left(\frac{\dot{r}}{r}\right)_{A,0} &= 2(1-u)H_0 \\ \left(\frac{\dot{\omega}}{\omega}\right)_{A,0} &= -3(1-u)H_0 \end{aligned} \quad (36)$$

An important property of equations (36) is that, for an $A$ observer, orbital characteristics (excluding tidal and other effects) depend only on $L$ (note that as $L = \alpha c$ it is $(\dot{L})_{R,0} = -(1-u)H_0$), being independent of light speed $c_R$.

### D. Rotation

In what concerns the rotational motion of an isolated solid body, its behavior is a consequence of the variation of the diameter, $d_R$, with $L$ and of the conservation of $J_R$. Representing the rotation angular velocity by $\Omega$ it is:

$$d_R = d_0 \alpha c$$
$$\frac{\Omega_R}{c_R} d_R^2 = \frac{\Omega_0}{c_0} d_0^2 \quad (37)$$

Thus $\Omega_R = \Omega_0 \alpha^{-2} c^{-1}$. In $A$

$$\left(\frac{\dot{\Omega}}{\Omega}\right)_{A,0} = (1-u)H_0 \quad (38)$$

The $A$ angular rotational velocity of an isolated solid body varies with $L^{-1}$, i.e., it increases if $L$ decreases through time.

### E. Moon-Earth-Sun system

To analyze the Moon-Earth-Sun system, the consequences of all tidal and other effects have to be considered. The equations for Moon orbit (39) and Earth rotation (40) are:

$$\left(\frac{\dot{r}}{r}\right)_{A,0} = 2(1-u)H_0 + \left(\frac{\dot{r}}{r}\right)_{Tidal}$$
$$\left(\frac{\dot{\omega}}{\omega}\right)_{A,0} = -3(1-u)H_0 + \left(\frac{\dot{\omega}}{\omega}\right)_{Tidal} \quad (39)$$

$$\left(\frac{\dot{\Omega}}{\Omega}\right)_{A,0} = (1-u)H_0 + \left(\frac{\dot{\Omega}}{\Omega}\right)_{Tidal} + \left(\frac{\dot{\Omega}}{\Omega}\right)_{Other} \quad (40)$$

It is long known that Moon is receding from Earth. The Moon-Earth distance increase calculated from lunar laser ranging is 3.8 cm y$^{-1}$, which corresponds, for $h_0 = 0.5$, to a rate of $1 \times 10^{-10}$ y$^{-1} = 2H_0$. Therefore, this distance increase corresponds to the first of equations (39) for $u = 0$. In this case, the consequence of tidal effect would be only marginal. Other interpretations of this result considering $u \neq 0$ imply that the tidal effect component on the increase of Moon-Earth distance be in the order of magnitude of $H_0$, in spite of depending on a variety of factors apparently unrelated to $H_0$. Note that although a large work has been made in the attempt of separating the variation in angular velocity due to the tidal effect from a variation with cosmological origin [23], following a suggestion from Hoyle, no conclusive results have been obtained.

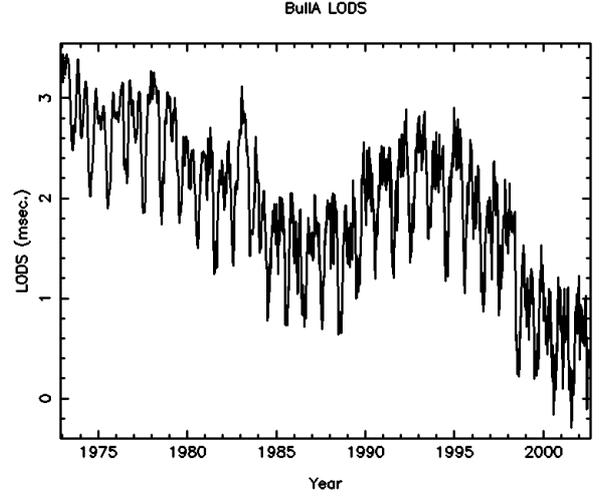

FIG. 3. – Excess to 86400 of the duration of the day, corrected to remove tides due to the solid Earth and oceans [24]

In what concerns Earth rotation rate in $A$ (40), a time variation of physical parameters adds a positive component for $u<1$, which corresponds to a variation in the length of day of $\Delta$LOD $= (u-1)H_0 \times$LOD $= -0.44$ ms cy$^{-1}$ for $u=0$ and $h_0=0.5$. This possibility is new. Note that methods like VLBI can directly determine $\Delta$LOD in $A$ while methods based on Earth past records or relative to orbiting bodies, either planets or satellites (GPS and others), do not directly determine $\Delta$LOD but are also sensitive to the variation of orbital motion. Several analyses of earth's rotation point out the existence of an acceleration of rotation with a non-tidal origin [25]. Instead of the expected value, from classical tidal explanation of Moon receding motion, of $\Delta$LOD = 2.3 ms cy$^{-1}$, the analysis of eclipses over the past 2500 years [26] obtained a value of only 1.8 ms cy$^{-1}$ and an analysis of a palaeoclimate record from the eastern Mediterranean Sea over the past three million years obtained 1.1 ms cy$^{-1}$ [27]. The last 300 years of analysis of LOD do not display any consistent increase [28], and the last 30 years of data (Figure 3), based on increasingly accurate space techniques, display an increase of earth rotation speed. These results are compatible with a tidal effect lower than it has been considered plus the above positive component, although other explanations are possible. The analysis of some cases of millisecond pulsars may be of interest.

### F. Solar irradiance

An important consequence of an orbital evolution is the variation of the solar flux received in each planet. The designation solar irradiance will be used for the solar flux measured by an *A* observer, $B_A$. As a result of the variation of sun-planet distance (35), the solar irradiance $B_A$, (neglecting the attenuation of radiation with time in the path between sun and planet) varies through time as

$$B_A = B_0 \frac{\ell_A(t)}{\ell_0} \alpha^4 c^4 \qquad (41)$$

In Figure 4 it is presented the variation of the solar irradiance on Earth considering the evolution through time of sun's luminosity $\ell_A(t)$ according to current stellar theory [29, 30] and the parameterization for $\alpha(t)$ and $c(t)$ used before. For $u=1$ an *A* observer cannot detect any evolutionary behavior at the solar system scale but will detect it at a cosmological scale. For $u=2/3$ and $h_0=0.5$ (or, in general, for $(1-u)h_0\approx 1/6$) the solar irradiance changed less than 1% in the last 4 Gy. For $u<2/3$, solar irradiance in Earth was greater in the past than today, attaining Venus present value at $4.2\times 10^9$ years ago for $u=0$ and $h_0=0.5$ (or $(1-u)h_0\approx 1/2$).

A past solar irradiance higher than today can be an explanation for the possible occurrence, at around 3.8 Gy ago, of large water flows / oceans in Mars [31, 32, 33, 34] and for the lost of large amounts of water since the end of hydrodynamic escape [35]. In what concerns the Earth, observational results seem to indicate a past climate warmer than present at most of the time, punctuated with several glaciations. The warmer periods are being currently explained considering changes in atmospheric composition influencing greenhouse effect. However, such possibility is not proved; on the contrary, there are evidences of solar influence on climate [36] suggesting that sun can be the determinant agent of large time scale climate variations. A convenient method of analysis is to establish the limit cases, i.e., a "cold earth" model (CEM) and a "hot earth" model (HEM). A CEM will not be very different from current models of earth past climate since these ones consider a solar irradiance correspondent to $u=1$. On the contrary, a HEM needs yet to be properly explored. A HEM can imply a very different understanding of several important observational evidences. For instance, the past low level of atmospheric oxygen can be explained as being mainly the result of a higher surface temperature, able of producing a large water vapor amount and, as a consequence, a lower atmospheric rate of oxygen. On the other hand, objections to a higher past temperature concerning life limits no longer stand because the phylogenetic evidence suggests that the ancestral bacteria were thermophiles [37]. The establishment of a HEM is of crucial importance for analyzing the past solar irradiance on earth.

## VIII. ON THE VARIATION OF PHYSICAL PARAMETERS

The positional dependence of matter/space properties is quite different of the dependence of atomic properties with motion. The latter phenomenon is an expectable (necessary?) consequence of field propagation speed being independent of source motion, while the former is not, its eventual cause being, at this point, unknown. However, one can note that to each atomic particle is associated an expanding field. The relation between particle and field is still unknown and different scenarios can be conceived, namely one where the particle is source or drain of the field. Such a scenario has an important set of consequences, one of them being time varying physical parameters. As far as this variation is locally undetectable by an atomic observer, i.e., obeying LRP, there is no *a priori* contradiction with experience. The analysis of particle-field relation is therefore crucial to understand the time variation of physical parameters.

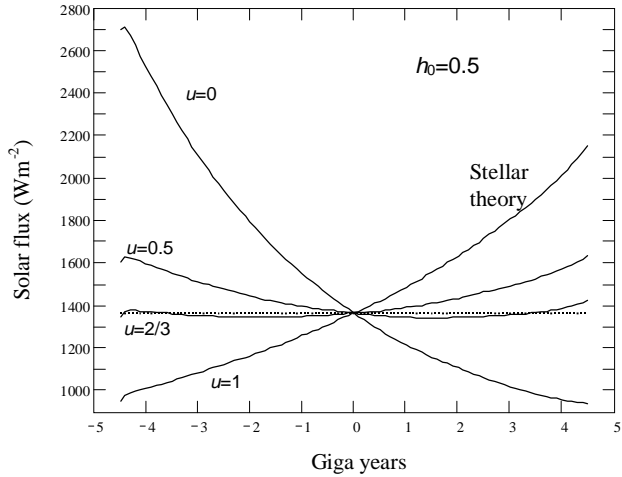

FIG. 4. Solar irradiance on Earth for $h_0=0.5$. The dotted line is the present value. The curve $u=1$ (invariance of *L*) corresponds to the variation of solar luminosity. For $u=2/3$ solar irradiance would have been almost constant in the past, while for $u=0.5$ solar irradiance will be almost constant in the next 4 Gy. For $u=0$ (invariance of light speed) solar irradiance would have been significantly higher in the past, attaining Venus present value at 4.2 Gy ago

## IX. CONCLUSION

Local Relativity establishes how matter/space characteristics can vary so that local physical laws hold. We have defined, in our previous paper, the fundamental framework of Local Relativity, composed of Local Relativity Property, LR conditions, an atomic and an invariant observer, a Euclidean space with no connection between time and space. We have now to determine the variation of physical parameters implied by the invariance of local physical laws, whatever the motion, field and position in space and time of the Einstein reference frame. In this paper we initiate the analysis of the positional case.

The first step of the analysis is the definition of a time varying scenario from redshift and CMB (cosmic microwave background). This means that we defined a positional variation of physical parameters liable to account for cosmic observations but in such a way that Relativity Principle is satisfied at least at local scale (what we called the Local Relativity Property).

An important result is that atomic time unit is time decreasing; a consequence is that in atomic units old stars and the Universe have approximately the same age while in invariant units matter is much older then stars. This is of utmost importance for large-scale structure analysis. Classic cosmic observational results are obtained, without any additional hypothesis, and compared with those of Friedmann models. Then, we analyzed the consequences at Solar System scale. This is a crucial point because no cosmological theory has ever succeed this test but by stating that does not apply at this scale. We concluded that, from the measures made in each time moment at Solar System scale, an atomic observer concludes that physical parameters are invariant when they vary accordingly with LR conditions. Therefore, the theory seems to fit both local and cosmic fundamental data. On the other hand, new results are also obtained concerning the evolution of Solar System. One, obtained for the first time, is the possibility of an accelerating component in Earth rotation motion, something that has been suspected in a number of analyses. This may be difficult to test due to the complexity of Earth rotation. The others are the receding motion of the Moon (and satellites) and a past higher temperature on earth (and other planets, as Mars). We show that, in our framework, there is no «a priori» incompatibility between these results and available data. The set of these consequences may support discriminative evidence.

## APPENDIX: ATTENUATION OF RADIATION

To analyze the power attenuation of radiation, i.e., CMB characteristics, we have to analyze a past Planck radiation. As Planck formula has no time derivative, we will consider it is valid both in $R$ and in $A$, considering $R$ or $A$ measuring units and constants, which can have different values. To analyze the present characteristics of a past Planck radiation, one has to express Planck formula in $R$ units, which are identical to present atomic units, and in function of present values of constants, which are the $A$ values. Noting that $h_R = h_A McL$, $T_R = T_A Mc^2$, $c_R = c_A LT^{-1}$ and $k_R = k_A$, it is:

$$dP_R(T_R, \lambda_R) = \frac{2\pi h_A c_A^2}{\lambda_R^5} \left[ \exp\left(\frac{h_A c_A L}{\lambda_R k_0 T_A}\right) - 1 \right]^{-1} d\lambda_R \cdot Mc^3 L$$
$$= Mc^3 L \cdot dP_A(T_A/L, \lambda_R)$$
(42)

To understand the meaning of (42) consider a gas in some distant past moment. To a local $A$ observer, the gas has a temperature $T_A$ and its thermal radiation, measured by $A$, matches the $A$ calculation using Planck formula for $T_A$. To a local $R$ observer, the gas temperature is $T_R = T_A Mc^2$ and its measures of radiation are in accordance with its calculation using Planck formula for $T_R$. Note that the measuring units and the values of constants are different for $R$ and $A$. Now consider that we measure that radiation, using our units, which have the value of the $R$ units, and using our values for constants, which are the $A$ values. We will conclude that our measures correspond to $Mc^3L$ times the Planck curve for $T_R/(Mc^2L) = T_A/L$, as (42) shows.

During its course, the wavelength of the radiation varies proportionally to $c_R$, thus the emitted wavelength is $c_R/c_0$ times the received one, i.e., $\lambda_R = c \cdot \lambda rec$. If there were no attenuation during the course, from (42) the received radiation, $dWrec$, would be:

$$dWrec = Mc^{-1}L \cdot dP_A[T_A/(z+1), \lambda rec]$$ (43)

As the CMB fits Planck formula, it is then necessary a flux attenuation of $Mc^{-1}L = M\alpha$ (in $R$). As the $A$ measuring unit of power flux varies with $M\alpha^{-3}$, the power flux attenuation in $A$ is $\alpha^4$, as postulated. Note that the apparent attenuation in $A$ is partly or totally due to the variation of $A$ measuring unit. For $M = \alpha^{-1}$, i.e., a time increasing of mass, there is no attenuation in $R$; for $M = 1$, i.e., a time invariant mass, the attenuation in $R$ is only $\alpha$.